# Realization of quasicrystalline quadrupole topological insulators in electrical circuits

Bo Lv[1†], Rui Chen[2,3,4†], Rujiang Li[5*], Chunying Guan[1], Bin Zhou[2], Guohua Dong[1], Chao Zhao[1], YiCheng Li[1], Ying Wang[1], Huibin Tao[6*], Jinhui Shi[1*], Dong-Hui Xu[2*]

[1]Key Laboratory of In-Fiber Integrated Optics of Ministry of Education, College of Physics and Optoelectronic Engineering, Harbin Engineering University, Harbin, China

[2]Department of Physics, Hubei University, Wuhan 430062, China

[3]Shenzhen Institute for Quantum Science and Engineering and Department of Physics, Southern University of Science and Technology (SUSTech), Shenzhen 518055, China

[4]School of Physics, Southeast University, Nanjing 211189, China

[5]Institute of Antennas and Electromagnetic Scattering, School of Electronic Engineering, Xidian University, Xi'an 710071, China

[6]School of Software Engineering, Xi'an Jiaotong University, Xi'an, China

[†]These authors contributed equally to this work.

*Correspondence to: rujiangli@xidian.edu.cn (R. Li), shijinhui@hrbeu.edu.cn (J. Shi) coldfire_arm@126.com (H. Tao), donghuixu@hubu.edu.cn (D.-H. Xu).

## Abstract

Quadrupole topological insulators are a new class of topological insulators with quantized quadrupole moments, which support protected gapless corner states. The experimental demonstrations of quadrupole-topological insulators were reported in a series of artificial materials, such as photonic crystals, acoustic crystals, and electrical circuits. In all these cases, the underlying structures have discrete translational symmetry and thus are periodic. Here we experimentally realize two-dimensional aperiodic-quasicrystalline quadrupole-topological insulators by constructing them in electrical circuits, and observe the spectrally and spatially localized corner modes. In measurement, the modes appear as topological boundary resonances in the corner impedance spectra. Additionally, we demonstrate the robustness of corner modes on the circuit. Our circuit design may be extended to study topological phases in higher-dimensional aperiodic structures.

**Introduction**

Since the discovery of topological insulators (TIs), tremendous effort has been devoted to the search for exotic topological phases of matter [1-10]. Recently, a novel class of exotic insulators with higher-order band topology, dubbed higher-order topological insualtors (HOTIs), have been proposed [11-25]. The fascinating characteristic of HOTIs is that a *d*-dimensional *n*th-order TIs have (*d*-*n*)-dimensional gapless boundary states [11-15]. For instance, two-dimensional (2D) second-order TIs with a quantized quadrupole moment, which is predicted by generalizing the fundamental relationship between the Berry phase and quantized polarization, support corner-localized modes lying in the middle of the energy gap [16]. Following the theoretical proposals of quadrupole topological insulators (QTIs), the topological corner states were observed experimentally in phononic [17-19] and photonic [20-25] systems.

The understanding of topological phases of matter has always been based on the topological band theory, which is defined in crystalline materials with long-range order and periodicity. To the best of our knowledge, topological phases are observed only in one-dimensional quasicrystals [26]. Although there are predictions of topological phases in 2D quasicrystals [27-38], the experimental observations have never been reported.

Highly customizable electrical circuits have emerged as a new platform to engineer various topological states [39-52]. In this paper, we report the realization of the 2D quasicrystalline QTIs in electrical circuits and observe the corner states experimentally. To realize the QTIs, we construct the Ammann-Beenker (AB) tiling circuits with two types (thin and fat) of rhombuses, and implement the nearest- and next-nearest-neighbor hoppings using lumped elements to realize rotational symmetry. Using impedance measurement, the corner states protected by quantized quadrupole moment are directly observed in the circuits. The robustness of topological corner states is also demonstrated. Our work provides the experimental evidence of topological phases in 2D quasicrystalline systems. We expect more topological states can be observed in circuits based on similar implementations.

**Results**

**Quasicrystalline quadrupole topological insulators**

We construct the QTI on an AB tiling quasicrystal by only considering nearest- and next-nearest-neighbor hoppings [34]. The tight-binding Hamiltonian is

$$H = A_0 \sum_j c_j^\dagger (\Gamma_2 + \Gamma_4) c_j + \frac{A_1}{2} \sum_{j \neq k} c_j^\dagger T(\phi_{jk}) c_k, \qquad (1)$$

where

$$T(\phi_{jk}) = \begin{cases} \Gamma_4 - i\Gamma_3, & -\pi/4 \leq \phi_{jk} < \pi/4, \\ \Gamma_2 - i\Gamma_1, & \pi/4 \leq \phi_{jk} < 3\pi/4, \\ \Gamma_4 + i\Gamma_3, & 3\pi/4 \leq \phi_{jk} < 5\pi/4, \\ \Gamma_2 + i\Gamma_1, & 5\pi/4 \leq \phi_{jk} < 7\pi/4, \end{cases} \qquad (2)$$

and $\phi_{jk}$ indicates the polar angle of bond between site $j$ and $k$ with respect to the horizontal direction. Here $c_j^\dagger = (c_{j1}^\dagger, c_{j2}^\dagger, c_{j3}^\dagger, c_{j4}^\dagger)$ is the creation operator in cell $j$. The first and second terms are the intracell and intercell hoppings with amplitudes $A_0$ and $A_1$, respectively. $\Gamma_4 = \tau_1 \tau_0$, and $\Gamma_\upsilon = -\tau_2 \tau_\upsilon$ with $\upsilon = 1,2,3$. $\tau_{1,2,3}$ are the Pauli matrices, and $\tau_0$ is the identity matrix. This model has four-fold rotational symmetry $\mathcal{C}_4 = UR$, where $U = \begin{pmatrix} 0 & i\tau_2 \\ \tau_0 & 0 \end{pmatrix}$ and $R$ is an orthogonal matrix permuting the sites of the tiling to rotate the whole system by $\pi/4$.

The rotational symmetry $\mathcal{C}_4$ results in a quantized quadrupole moment $Q_{xy} = 0, e/2$. Thus, $Q_{xy}$ is a natural topological invariant, which can be calculated in real space [53,54]. By numerical calculations, we confirmed that the AB tiling quasicrystal has the non-trivial quadrupole moment $Q_{xy} = e/2$ under the hopping ratio $\lambda > 2.5$ ($\lambda = A_1/A_0$) (See Supplementary Note 1). The quantized quadrupole moment indicates the occurrence of the quadrupole insulator with protected corner states.

**Realization in electrical circuits**

The quasicrystalline lattice can be mapped to an electrical circuit. To realize quasicrystalline QTI, we design an electrical circuit depicted in Figure. 1a. The circuit is characterized by the Kirchhoff's law

$$I_{pa} = \sum_{qb} J_{pa,qb}(\omega) V_{qb}(\omega), \qquad (3)$$

where $I_{pa}(t) = I_{pa}(\omega)e^{i\omega t}$ [$V_{qb}(t) = V_{qb}(\omega)e^{i\omega t}$] is the current (voltage) at site $a$ ($b$) in cell $p(q)$, and $\omega$ is the frequency of the circuit. The circuit Laplacian is $J_{pa,qb}(\omega) = i\omega H_{pa,qb}(\omega)$, where

$$H_{pa,qb}(\omega) = C_{pa,qb} - \frac{1}{\omega^2} W_{pa,qb}, \qquad (4)$$

with $C_{pa,qb}$ being the capacitance between two sites, and $W_{pa,qb} = L^{-1}_{pa,qb}$ being the inverse inductivity between two sites. Here the subscript $g$ means the ground. $H_{pa,qb}$ can be equivalent to the Hamiltonian of quasicrystal in Eq. (1) if the diagonal elements are zero at $\omega_0$. For the diagonal components with $pa = qb$, the grounded elements are chosen for satisfying $C_{pa,pa} = -C_{pa,g} - \sum_{q'b'} C_{pa,q'b'}$ and $W_{pa,pa} = -L^{-1}_{pa,g} - \sum_{q'b'} L^{-1}_{pa,q'b'}$.

As shown in Figure. 1a, the solid and dashed lines correspond to hoppings with positive and negative values, which can be realized by the capacitors and inductors, respectively. We choose capacitor $C_1$ and inductor $L_1$ for intracell hopping, and $C_2$ and $L_2$ for intercell hopping. The nearest neighbor couplings between cells are introduced to stabilize the corner states in small size systems [34]. If these couplings is neglected, the exact AB tiling is restored, and the physical results don't change qualitatively.

Under a suitable choice of the grounded elements, the tight-binding Hamiltonian of the QTI on a quasicrystalline lattice in Eq. 1 is mapped to the circuit Hamiltonian in Eq. 2. Note that there exists a relation $C_2/C_1 = L_1/L_2 = A_1/A_0$.

The two-point impedance is:

$$Z_{ab} = \frac{V_a - V_b}{I_{ab}} = \sum_n \frac{|\psi_{n,a} - \psi_{n,b}|^2}{j_n} \quad (5)$$

where $V_{a(b)}$ is the voltage on site $a(b)$, $I_{ab}$ is the current between the two sites, and $j_n$ and $|\psi_n\rangle$ are the eigenvalue and eigenstates of the matrix $J(\omega)$, respectively [21]. The two-point impedanace $Z_{ab}$ diverges in the presence of zero-admittance modes ($j_n = 0$) with $\psi_{n,a} \neq \psi_{n,b}$. Hence, the corner states with zero-admittance can be observed by measuring the two-point impedance.

We consider the QTI on an AB tiling 2D quasicrystalline lattice. A circuit that realizes the QTIs is depicted in Figure. 1a. The unit cell of the circuit contains four sites denoted by labels 1, 2, 3, 4. Each unit cell consists of three capacitors for positive coupling and one inductor for negative coupling, which can generate a synthetic magnetic π flux threading the unit-cell plaquette (equivalent to half the magnetic flux quantum, $(1/2)\Phi_0 = h/(2e)$, where $h$ is the Planck constant). The existence of this non-zero flux opens the spectral gap for maintaining the corner-localized mid-gap modes. We use two pairs of capacitors and inductors ($C_1$, $L_1$) and ($C_2$, $L_2$), which have the same resonant frequency $\omega_0 = 1/\sqrt{L_1 C_1} = 1/\sqrt{L_2 C_2}$, as the intracell and intercell wirings between the sites, respectively. The intracell and intercell elements are related by $C_2 = \lambda C_1$ and $L_2 = L_1/\lambda$. The circuit is governed by the linear circuit theory with a circuit Laplacian $J(\omega)$ [21]. The circuit has a square-open boundary which satisfies the global $\mathcal{C}_4$ symmetry at the resonant frequency $\omega_0$. Suitable grounded elements on the sites are chosen (site colors depicted in Figure. 1a indicate the values of the grounded capacitors and/or inductors) to sustain chiral symmetry at $\omega_0$, which pins the topological boundary modes in the middle of the bulk energy gap. The symmetry characteristic and quantized quadrupole moment of circuit indicate the occurrence of the QTI associated with topologically protected states localized on the corner sites. The edge states are gapped and merge with the bulk states. Hence it is difficult to observe them experimentally (See

Supplementary Note 2).

To realize the QTI experimentally, a circuit with 69 unit cells was fabricated as shown in Figure. 1b. The intracell elements with $C_1$ = 10 nF and $L_1$ = 1 mH result in a resonant frequency 50.3 kHz (parameters for other elements are given in Methods). Here we set the coupling ratio $\lambda = 10$ in order to get highly localized corner states, which can be observed by two-point impedance measurements between the corner/edge/bulk site and another bulk site using an impedance analyzer.

Figure 2 compares the experimental and theoretical results, which demonstrates the spectral and spatial localizations of the topological corner states. Figure. 2a shows the spectrum of the circuit Laplacian $J(\omega)$ as a function of the normalized frequency $\omega/\omega_0$. The isolated corner modes reside in the spectral gap of $J(\omega)$ at a fixed frequency $\omega_0$. The spatial distribution of the experimentally observed impedance of the corner states at $\omega_0$ is illustrated in Figure. 2b. The impedance is maximum at the four corner sites and exponentially decays at other sites. The comparison between the experimental and theoretical impedance spectra is shown in Figures. 2c, d, which demonstrate the spectral localization of the corner modes. The maximum measured impedance reaches 7 kΩ.

To justify the robustness of the corner states, we consider the effect of experimental element errors. The errors are generated in two ways: (i) random manufacturing variations in discrete elements, and (ii) the deviation of the typical values of commercially available circuit components from the theoretical circuit parameters. Figure. 3a shows the normalized eigenfrequencies $\omega/\omega_0$ of the Laplacian $J(\omega)$ with element error ±5%, and the number of the samples is 300. The eigenfrequencies of corner states (indicated by red circles) are located in the bandgap and far away from those of other states (indicated by blue circles). Experimentally, we construct a circuit with element error ±5%, and measure the two-point impedances as a function of the normalized frequency $\omega/\omega_0$. As shown in Figure. 3b, the corner states are spectrally localized with a deviation from $\omega_0$. Figures. 3c-f show the spatial impedance distributions at the frequencies of the four corner states. Compared with the spatial impedance distribution in Figure. 2b, although the maximum impedances are located at only one of the four corner sites, the topological corner states are still spatially localized. The spectral and spatial localizations imply the robustness of corner states on the circuit.

**Discussion**

The quadruple insulators were always realized on periodic systems with translational symmetry, and the quadruple moments are well defined in momentum space. Unlike the periodic structures, the quasi-periodic systems possess long-range order but don't have translational symmetry. Therefore, quasicrystalline insulators with a quantized quadrupole moment extend the concept of quadruple insulators defined in crystals. The realization of quadruple isolator in this work opens a way for the implementation of corner states in more quasi-periodic systems.

The circuit implementation of the 2D quasicrystalline QTIs confirms the existence and robustness of the corner states on the circuit. This work provides an experimental evidence for the first time to implement the topological phase in a 2D quasicrystalline system, and extends the territory of topological phases beyond crystals to higher-dimensional aperiodic systems. The highly customizable circuit platform can also readily be applied to other 2D quasicrystals with different symmetries and to 3D quasicrystalline structures [55], which may realize more exotic topological states.

**Acknowledgements**

B.L. was sponsored by the Fundamental Research Funds for the Central Universities under Grant No. 3072020CFT2501 and 3072020CF2528, the National Natural Science Foundation of China under Grant No. 61901133. R.C. acknowledges support from the project funded by the China Postdoctoral Science Foundation (Grant No.2019M661678). J.S. was sponsored by the National Natural Science Foundation of China under Grant Nos. 91750107 and 61875044. D.-H.X. was supported by the NSFC (Grant No. 11704106). D.-H.X. also acknowledges the financial support of the Chutian Scholars Program in Hubei Province.

**Methods**

We choose the capacitors $C_1$ = 10 nF, $C_2$ = 100 nF, $C_{1g}$ = 100 nF, $C_{2g}$ = 200 nF and the inductors $L_1$ = 1000 μH, $L_2$ = 100 μH with high $Q$ factors (Q > 70 @ 50 kHz). The grounded elements are $L_{1g}$ = 83 μH, $L_{2g}$ = 100 μH, $L_{3g}$ = 45 μH, $L_{4g}$ = 31 μH, $L_{5g}$ = 23 μH, $L_{6g}$ = 500 μH, and $L_{7g}$ = 50 μH. After delicately choosing, the errors of the circuit elements are smaller than ±1%. The intracell elements $C_1$ and $L_1$ result in a resonant frequency 50.3 kHz. Here we set the coupling ratio $\lambda$ = 10 in order to get highly localized corner states, and the two-point impedance measurement is carried out using an Impedance Analyzer 4192A LF.

**Data availability**

The data that support the findings of this study are available from the corresponding author upon reasonable request.

**Conflict of interest**

The authors declare that they have no conflict of interest.

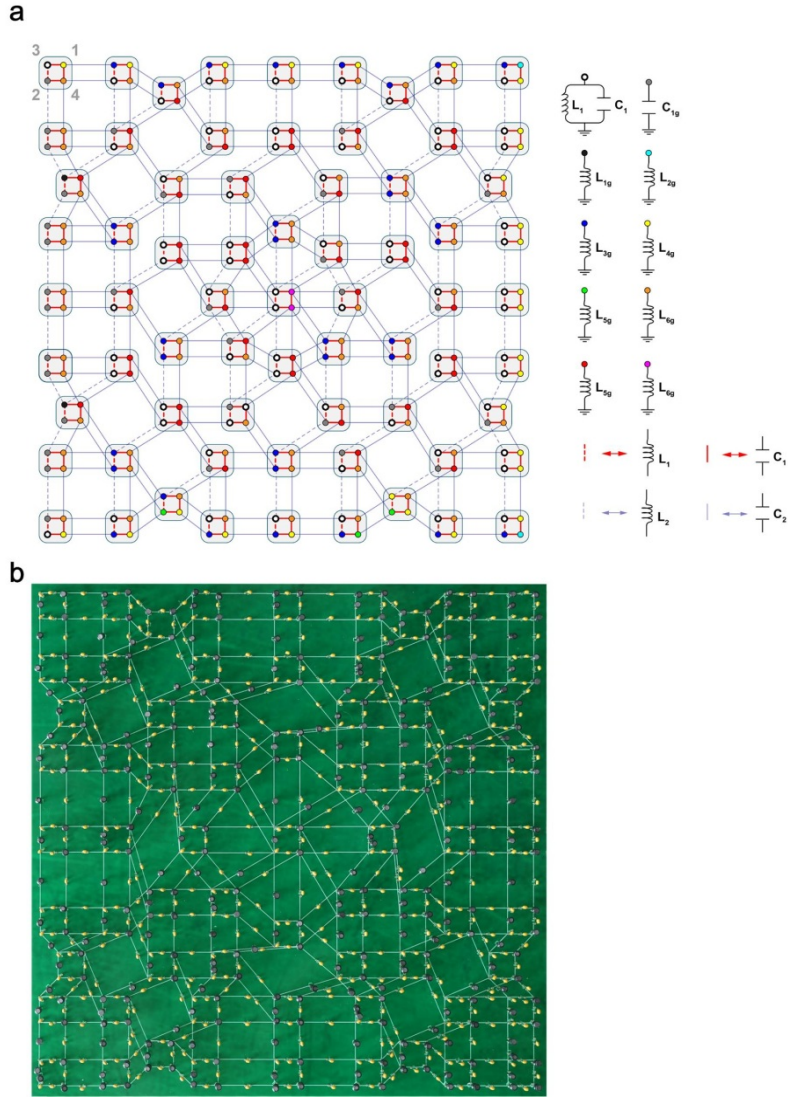

**Figure. 1 Quasicrystalline quadrupole topological insulators in electrical circuits.**
**a)** Schematic of the circuit with 69 unit cells. One unit cell consists of four sites labeled as 1, 2, 3 and 4 at the left-up corner. The colored sites are wired to grounded inductors and/or capacitors listed on the right. For example, the black hollow point 3 is connected to the ground by the capacitor-inductor pairs $C_1$ and $L_1$, and the yellow point 1 is connected to the ground by $L_{4g}$. The sites in the same unit cell are wired by intracell inductors $L_1$ (red dashed lines) and capacitors $C_1$ (red solid lines), and the intercell elements between the unit cells are inductors $L_2$ (gray dashed lines) and capacitors $C_2$ (gray solid lines). **b)** Photo of the fabricated circuit. The yellow and black elements are capacitors and inductors, respectively, and the wirings are indicated on overlay of the printed circuit board.

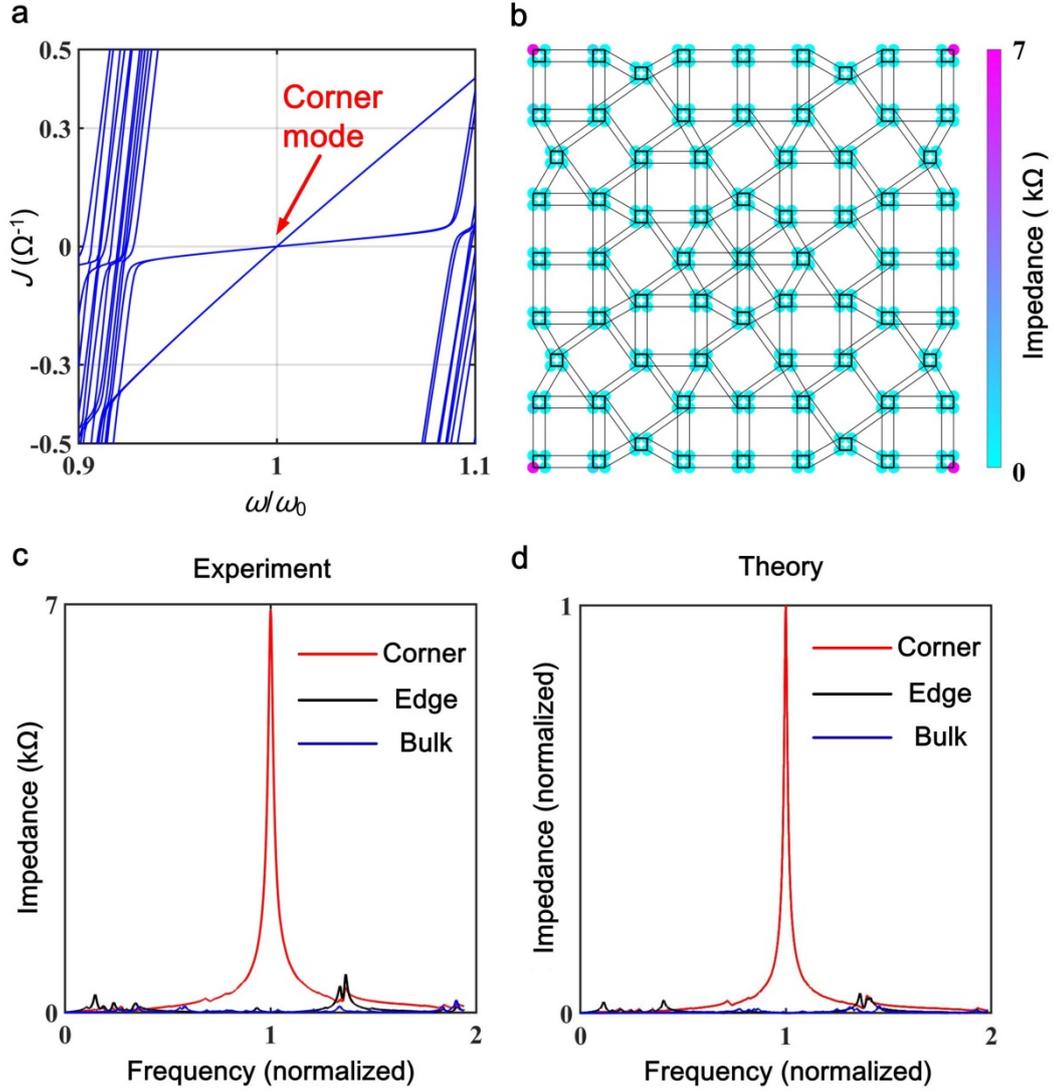

**Figure. 2 Corner states on the circuit. a)** Spectra of the Laplacian *J* versus the frequency which is normalized to the resonant frequency $\omega_0$. The isolated mode crossing in the gap corresponds to the topological corner states. **b)** The experimental impedance distribution at the resonant frequency $\omega_0$ demonstrates the spatial localization of the corner states. **c)** The experimental two-point impedances measured between the left-up corner/edge/bulk site and another bulk site. **d)** The theoretical two-point impedances. Both the experimental and theoretical results demonstrate the spectral localization of the corner states.

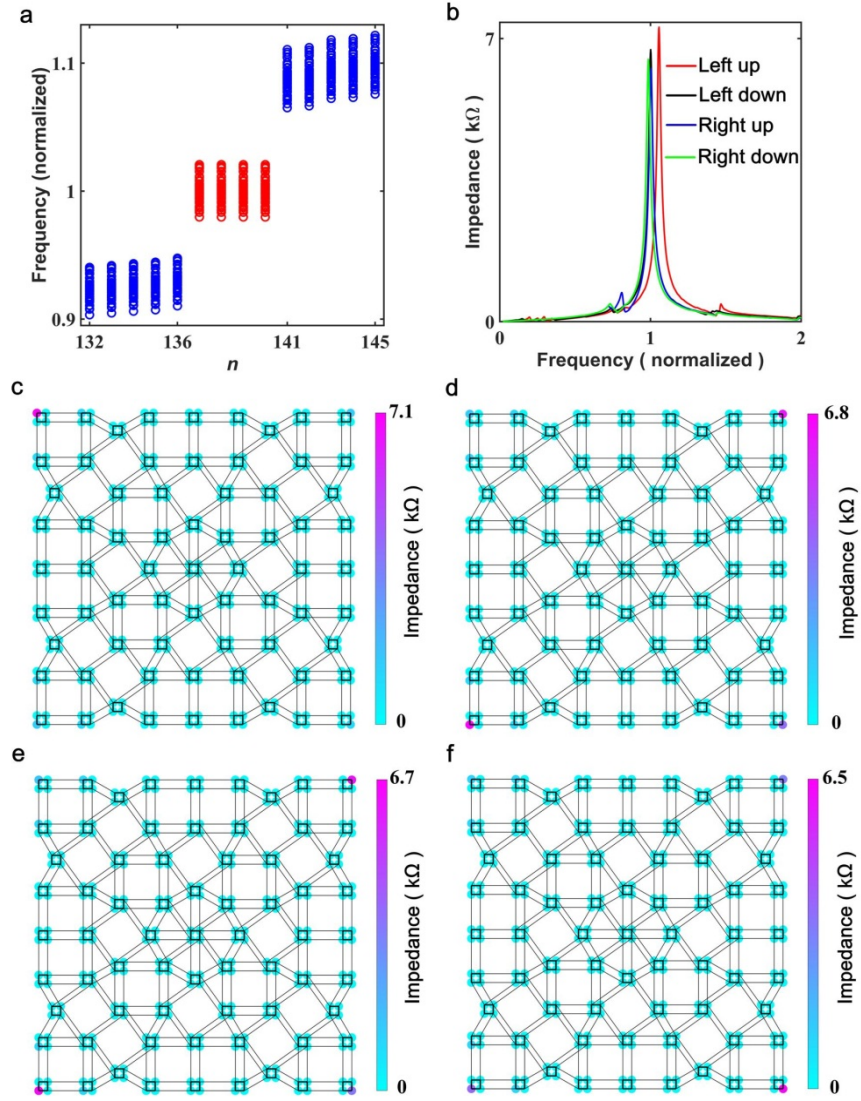

**Figure. 3 Robustness of corner states on the circuit. a)** The normalized eigenfrequencies $\omega/\omega_0$ of the circuit Laplacian $J(\omega)$ with element errors ±5%, and the number of the samples is 300. The eigenfrequencies of the corner and bulk states are indicated by the red and blue circles. **b)** The experimental two-point impedances measured between the left-up/left-down/right-up/right-down corner site and another bulk site. **c-f)** The experimental impedance distributions at the frequencies of the left-up/left-down/right-up/right-down corner states.